# Commenter Behavior Characterization on YouTube Channels

Shadi Shajari
COSMOS Research Center
University of Arkansas at Little Rock
Little Rock, USA
Email: sshajari@ualr.edu

Nitin Agarwal
COSMOS Research Center
University of Arkansas at Little Rock
Little Rock, USA
Email: nxagarwal@ualr.edu

Mustafa Alassad
COSMOS Research Center
University of Arkansas at Little Rock
Little Rock, USA
Email: mmalassad@ualr.edu

*Abstract*—YouTube is the second most visited website in the world and receives comments from millions of commenters daily. The comments section acts as a space for discussions among commenters, but it could also be a breeding ground for problematic behavior. In particular, the presence of suspicious commenters who engage in activities that deviate from the norms of constructive and respectful discourse can negatively impact the community and the quality of the online experience. This paper presents a social network analysis-based methodology for detecting commenter mobs on YouTube. These mobs of commenters collaborate to boost engagement on certain videos. The method provides a way to characterize channels based on the level of suspicious commenter behavior and detect coordination among channels. To evaluate our model, we analyzed 20 YouTube channels, 7,782 videos, 294,199 commenters, and 596,982 comments that propagated false views about the U.S. Military. The analysis concluded with evidence of commenter mob activities, possible coordinated suspicious behavior on the channels, and an explanation of the behavior of co-commenter communities.

*Keywords-Social Network Analysis*; *YouTube*; *Commenter Network Analysis*; *Principal Component Analysis*; *Suspicious Behaviors*;

## I. INTRODUCTION

YouTube is a popular online platform for sharing and discussing videos, which allows millions of users to upload content and comment on videos daily. While YouTube has many benevolent uses, it has also been used to spread misinformation, propaganda, and other inappropriate or malicious content [1]. In addition, the comments section on YouTube has been used as a breeding ground for suspicious commenter behavior. For example, adversarial information actors deployed an information tactic known as commenter mobs, a strategy of random group of commenters collectively comment on a video (or a set of videos) to boost the video's engagement (hence, fabricate virality), as investigated by Hussain et al. [2].

Moreover, such mobs of commenters could comment on videos on one or multiple channels, where these comments may or may not be relevant to the videos. This behavior can harm the quality of the YouTube community and shape public perceptions of important issues.

Prior studies on this issue have mostly focused on identifying certain keywords and phrases that may indicate suspicious behavior. Researchers applied methods to analyze the behavior of the key sets of commenters in co-commenter networks on YouTube [3]. However, we discovered some limitations in these methods as well as shortcomings in the final analysis, in which the authors did not count the suspicious behavior of commenters at the channel level.

This paper proposes a social network analysis based methodology to detect commenter mobs, which helps to develop a channel-level characterization from highly suspicious to least suspicious level. In addition, the methodology helps assess the similarity between channels based on commenter behaviors, which allows the detection of varying degrees of collusion (or coordination) among channels. This systematic analysis was implemented to investigate suspicious commenter behavior on 20 YouTube channels promoting false views of the U.S. military and claiming to report official news and information about the U.S. Department of Defense. Likewise, mob activities were detected in this collection of channels. For example, the method found activities in standalone channels that shared similar organizational structures (mob leaders and affiliates), comment posting styles, and languages. Further analysis was done on the commenter mobs of the three most suspicious channels to describe their similarities and the nature of the content they posted. This study has contributed significantly and generated new insights into analyzing inorganic behaviors in YouTube platform, as presented below:

- Conducted an analysis of co-commenter networks on 20 YouTube channels to identify similarities using clustering methods and Principal Component Analysis (PCA) for dimensionality reduction.
- Detected the commenter mobs of the three most suspicious channels.
- Investigated cross-channel activities among the three most suspicious channels.

The rest of the paper is organized as follows: Section II reviews the suspicious behavior on YouTube, commenter network analysis, and the current understanding of the topic. Section III details the methods used for data collection, including the techniques and tools utilized to collect data. Section IV explains the methodology used in the study, which involves a combination of PCA, k-means, and hierarchical clustering methods. Section V presents the results of the study, including a detailed analysis of the data, commenters behavior, and the YouTube channels analysis. Finally, Section VI provides a conclusion and recommendations for further research.

## II. RELATED WORK

In this section, we review the relevant literature related to suspicious behaviors on YouTube. Since this area is vastly understudied, we expand our literature survey to include approaches that are methodologically relevant to ours, even if they are studying a different research problem.

*A. Suspicious Behavior on YouTube*

Several approaches have been proposed to study suspicious behaviors on YouTube. A study by Alassad et al. [1] discovered intensive groups on YouTube, and identified content-user networks that responsible for disseminating conspiracy theories using a two-level decomposition optimization method. Likewise, Alassad et al. [3] applied the bi-level max-max optimization approach to identify key sets of individuals on social media who have the power to mobilize crowds and regulate the flow of information. Research by Kaushal et al. [4] focused on detecting child unsafe content on YouTube, using supervised classification and a convolutional neural network with an accuracy of 85.7%. Furthermore, Hussain et al. [2] analyzed metadata, such as engagement scores to identify inorganic behaviors on YouTube.

Kirdemir et al. [5] presented an unsupervised model for co-ordinated inauthentic behavior assessment on YouTube using a methodology that combines multiple layers of analysis, including rolling window correlation analysis, anomaly detection, peak detection, rule-based supervised classification, network feature engineering, and unsupervised clustering approaches. These studies come close to analyzing suspicious behaviors on YouTube. There is still a gap in the literature on examining suspicious commenter behaviors on YouTube that may lead to engagement boosting or fabricating the virality of the content. We focus on addressing this knowledge gap at the channel level through the research presented in this paper.

*B. Commenter Network Analysis*

An extensive body of literature applies social network analysis methods to reduce the complexity of social media data structures. Reviewing all those studies would be outside the scope of this research. Therefore, we narrow our focus on the studies that utilize social network analysis to analyze networks on YouTube only. Shapiro et al. [6] analyzed the video-commenter discussion on YouTube. They presented a comprehensive network analysis for the 20 most popular climate change-related YouTube videos to understand the role of elites and small groups of frequent commenters in shaping the discussion.

More recently, studies have focused on exponential random graph modeling [7]. Another study by Ferreira et al. [8] examined the mechanisms of imitation, intergroup interaction, and communities of co-commenters. In this research, the model focused on groups of users who frequently interact by commenting on the same posts. Coppola et al. [9] found that machine learning can detect maximal cliques within larger networks by implementing the Bron-Kerbosch algorithm to detect communities and central nodes. Likewise, the study presented by Cascavilla et al. [10] focused on analyzing commenters' comments and social network graphs to identify censored individuals in news articles. Another study by Wattenhofer et al. [11] analyzed the social and content aspects of YouTube and compared them to traditional online social networks, such as Twitter.

However, there are gaps in the literature related to commenter behavior on YouTube that could be addressed in future research. For example, we noticed a need for more research on the effectiveness of different approaches to detecting and addressing the multi-channel suspicious commenter behavior. In this research, we propose graph theoretic methods to identify mobs of commenters who post comments together on one or a set of videos and exhibit similar behaviors on different Youtube channels, in an attempt to discover collusion (or coordination) among channels.

### III. DATA COLLECTION

In this study, we used a Python-based multi-thread script [12] to collect data related to the U.S. Military on YouTube channels. For this purpose, subject matter experts identified a list of 20 YouTube channels that promoted false views of the U.S. military.

TABLE I. YOUTUBE DATASET STATISTICS

| Channels | Videos | Comments | Commenters |
|---|---|---|---|
| 20 | 7782 | 596,982 | 294,199 |

The YouTube Data API is utilized to retrieve huge amounts of data about the number of videos, the number of comments, the details of the comments, and the commenter's IDs as presented in Table I.

### IV. METHODOLOGY

This section describes our model. First, we describe how to create a co-commenter network (Section IV-A) and then discuss the 20 network structural features extracted from the co-commenter network (Section IV-B). Next, the model utilizes k-means and hierarchical clustering methods [13] on the co-commenter networks to determine the level of similarity between all 20 YouTube channels. The network structural features also allow us to rank the suspiciousness of the channels based on their co-commenting behaviors. Next, we explain the model in detail.

*A. Creating Co-commenter Network*

The process begins by creating a co-commenter network for each channel based on a chosen threshold. This co-commenter network is made up of edges between commenters who have commented on the same video, with the weight of the edge representing the number of the same videos they have commented on. Let's call this number $n$. To maintain stronger co-commenter relations, a threshold needs to be identified for $n$. To find the optimal threshold, 10 co-commenter networks were created for each channel using thresholds from 1 to 10. The average clustering coefficient was calculated for each network at each threshold, and the results were analyzed. The optimal threshold for each channel was determined by identifying the point in the plot where the rate of change starts to decrease, which is commonly known as the "elbow point" [14]. This method is often used to determine the optimal number of clusters in a data set. We analyze the co-commenter network with the best threshold that is calculated through the "elbow point" for each channel.

*B. Extracting Co-commenter Network Features*

Kirdemir et al. [5] have developed a method to detect suspicious clusters and behavior on YouTube channels by analyzing the interactions between commenters. They established a set of network features, including metrics from well-established graph measures to analyze the networks' behavior from different dimensions. In our analysis, we utilized 20 network structural features to study the co-commenter networks, such as number of nodes, number of edges, total number of unique commenters, total number of comments, normalized ratio of co-commenters (nodes / total commenters), average degree, density, average clustering coefficient, modularity, number of maximal cliques that have at least 5 members, number of unique commenters in cliques, number of commenters in cliques / total number of commenters in the channel, number of commenters in cliques / number of nodes, average degree of cliques, average degree of cliques / average degree in the co-commenter network, average clustering coefficient of cliques, average clustering coefficient of cliques / average clustering coefficient in the co-commenter network, mean clique size, median clique size and maximum clique size.

The methodology presented in this paper attempts to discover the similarities between channels across all the features in the dataset using unsupervised methods, such as k-means and hierarchical clustering [15]. In other words, utilizing k-means and hierarchical clustering methods, we gain insights into the similarities of commenter behaviors exhibited in different channels. In addition, the Principal Component Analysis (PCA) is utilized to reduce the high-dimensional co-commenter network feature space, thereby reducing the complexity of the dataset, retaining important links between commenters, increasing the interpretability of the identified patterns, and minimizing information loss [16].

Additionally, channels are ranked on suspiciousness based on the highest number of commenters, comments, and maximal cliques that have at least five members (for having significant number of cliques). The behavior of the commenters across the top suspicious channels is analyzed. While per channel analysis allows us to see commenter mob behaviors on a particular channel, combined-channel analysis shows how commenter mobs span across multiple channels. Such an analysis shows how some influential commenter mobs move from one channel to another in an attempt to amplify/boost the videos' engagement in a highly sophisticated manner, much like a well-choreographed flash mob.

As the next step, social network analysis is conducted on the most suspicious channels. This analysis is done in order to gain insight into the behavior and interactions of those who comment on these three channels. Furthermore, the co-commenter network for each channel is examined separately, and communities are identified based on modularity methods. The influential commenters who posted the most comments within these communities were identified by analyzing the average degree centrality of nodes. The next section discusses our results and findings.

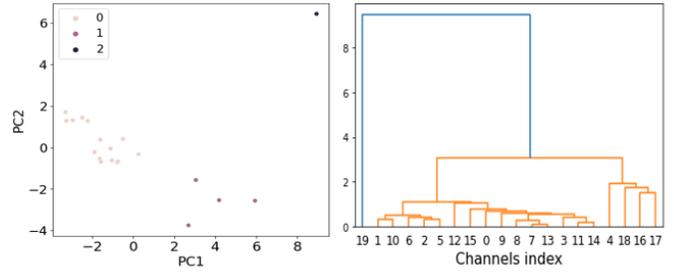

Figure 1. Channel categorization using k-means (left) and hierarchical clustering (right).

## V. RESULTS

This section will cover the results of K-Means and Hierarchical Clustering, followed by an examination of the findings from the analysis of commenter mobs on the top three suspicious channels.

*A. K-Means and Hierarchical Clustering*

This section discusses the results of clustering analyses performed using both the k-means and hierarchical clustering methods [13]. The utilized techniques identified three clusters in both methods. The optimal number of clusters for k-means was found by utilizing the silhouette score as described in [17], where a higher score indicates higher similarity among channels within a cluster. For hierarchical clustering as presented in [18], the single linkage method is used to create a dendrogram (tree-like structure) that shows how close each channel is to other channels.

Moreover, the cut-tree method is used to determine the optimal number of clusters and separate the channels into groups based on their similarity. Likewise, the optimal number of clusters helps to evaluate the quality of clustering by determining how similar an object is to its own cluster compared to other clusters. The visualizations of the clusters, such as scatter plots [19] demonstrated that the clusters were well separated. In other words, such a clustering method offers a better visual representation of how well each data point has been classified within a cluster. The clusters identified by both k-means and hierarchical clustering were distinct and did not overlap, as shown in Figure 1. The PC1 and PC2 correspond to the two principal components identified using the PCA method for dimensionality reduction, as explained in Section IV.B. The channel index is the unique channel ID assigned to each channel in our dataset. Both methods successfully identified the same number of clusters in the data with good separation between them. Finally, we examined the structure of clusters' networks based on three primary characteristics of the co-commenter network. In other words, we measured the average clustering coefficient (ACC), the modularity values, and the density to identify the characteristics of channels within these clusters, as shown in Table II.

In summary, the clustering analysis indicates that three channels require further investigation due to the network structures. Channels like "USA Military Channel" and "USA Military Channel 2" are classified as part of cluster 1, which is logical as they have similar video content and denser co-commentor network structures. On the other hand, "The Military TV" channel is part of cluster 0, which comprises channels with smaller co-commenter networks and distinct content.

### B. Commenter Mobs for All Three Channels

As described in Section IV-B, our model can rank channels' suspiciousness based on the highest number of commenters, comments, and maximal cliques that have at least five members. We identified the top three suspicious channels from our model for further analysis. The three most suspicious channels are "USA Military Channel" (1.53M subscribers), "The Military TV" (399K subscribers), and "USA Military Channel 2" (372K subscribers). We created a combined-channel commenter network for the three channels. Figure 2, illustrates groups of commenters across the three most suspicious channels. The structure of each channel is represented by a central node called the channel's name, and nodes connected to that channel are the commenters who post comments on that channel's videos. Next, we employed the modularity method to identify patterns and communities (different colors) within this network. The green group is "The Military TV" channel with ID "UC6qHUwB5F_fPPaRXL8COqew", the purple group is "USA Military Channel" with ID "UC5bu8qvD-y0eo2A6yLfvr3A", and the orange group is "USA Military Channel 2" with ID "UCpU184Ub1cuTsrHqb1RodjA". The channels' analysis tells us that the most active commenters are those who frequently post comments on both "USA Military Channel" and "USA Military Channel 2".

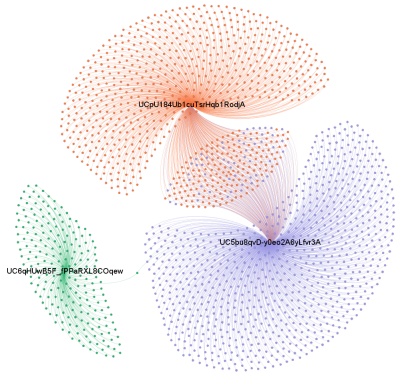

Figure 2. Commenter mobs spanning across the three most suspicious channels.

Moreover, these commenters show interest in channels that primarily post videos related to the United States military and other narratives, such as the Army, Navy, Air Force, and Marine Corps. From our analysis, we identified many Japanese comments posted by these commenters due to possible common interests in Japan's self-defense forces and NATO countries [20]. Moreover, Figure 2 shows the bridge commenters sitting at the intersection of two channels, "USA Military Channel" and "USA Military Channel 2". We could call such behavior "YouTube cross-channel activities" due to the behavior of these commenters. Finally, "The Military TV" channel offers viewers footage from various military branches, including the narrative weapons, aircraft, tanks, ships, guns, artillery, vehicles, military operations, and technologies of the US and other countries [21]. However, it appears that the commenters on this channel are less active than those on the other two suspicious channels. Next, we illustrate each channel's structure and discuss commenters' behavior.

### C. Commenter Mobs for USA Military Channel

In Figure 3, the co-commenter network presents 1.53M subscribers, and 41,092 commenters spread information across 1,993 videos. According to the modularity measurements, the co-commenter network in the USA Military Channel is divided into four distinct communities. Due to the space limit, the analysis focused on the top three largest communities shown in Figure 3. Moreover, the structures of communities (B), (C), and (D) indicate leaders and followers in organizations based on the degree centrality values. In other words, we observed commenters supporting comments posted by highly central commenters in the channel. For example, the leader-follower case was related to comments on videos related to the United States military, such as the Army, Navy, Air Force, and Marine Corps. In addition, our analysis shows that US allies were involved in many comments with Japanese contexts and videos related to the Japanese Self-Defense Forces and NATO countries.

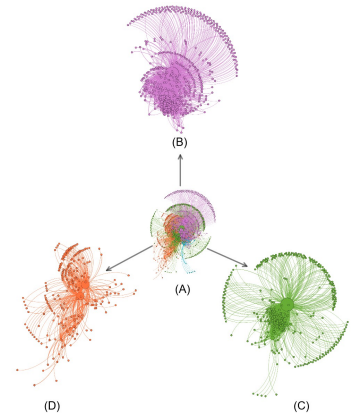

Figure 3. USA Military Channel's co-commenter network (A) and the three largest commenter mobs (B, C, and D).

### D. Commenter Mobs for USA Military Channel 2

Figure 4 illustrates the "USA Military Channel 2" which has 372,000 subscribers, and 37,527 commenters across 227 videos. The analysis shows similar behavior from the commenter as we implemented the exact steps explained in Section V-C. From our results, we observed similar context, language, and the same commenters were active on both channels. Likewise, these two channels display a strong resemblance in organizational structures on YouTube, indicating a potential level of collusion or coordination among commenters. Moreover, our analysis revealed commenters were active on both channels, "USA Military Channel" and "USA Military Channel 2" which indicates cross-channel activities on YouTube. Furthermore, these two channels were part of cluster 1 using k-means and hierarchical clustering methods, as shown in Figure 1.

TABLE II. CLUSTER STATISTICS AND DESCRIPTION

| Cluster | Acc | Density | Modularity | Description |
|---|---|---|---|---|
| 0 | 0.35 | 0.06 | 0.41 | Channels in this cluster have fewer network structures than the channels in other clusters. The co-commenter networks do not have many tightly knit groups of commenters and not cohesive structure. |
| 1 | 0.75 | 0.07 | 0.25 | Channels' co-commenter networks are well-connected networks with strong internal connections among commenters. |
| 2 | 0.52 | 0.002 | 0.24 | The clusters' channels show larger co-commenter network structures than the networks in other clusters. |

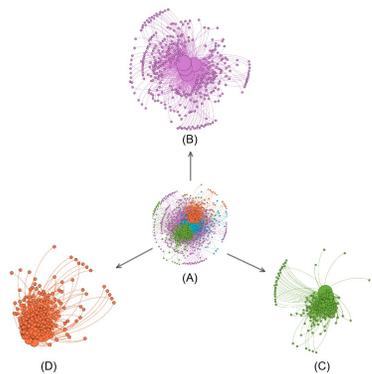

Figure 4. USA Military Channel 2's co-commenter network (A) and the three largest commenter mobs (B, C, and D).

*E. Commenter Mobs for The Military TV*

Figure 5 depicts the last suspicious groups of commenters in the structure of channel "The Military TV" This YouTube channel includes 399,000 subscribers, 54,898 comments, and 29,113 commenters across 582 videos. Our findings indicate that the network structure for this channel is smaller in terms of nodes, edges and the number of commenters in comparison to the other two channels. Additionally, the organizational structure of the channel is different from "USA Military Channel" and "USA Military Channel 2" similarly, k-means and hierarchical clustering methods put this channel in cluster 0 as shown in Figure 1. Likewise, our analysis showed different context used by commenters in this channel, where the videos and comments were related to the US and other countries, such as Russia and Ukraine.

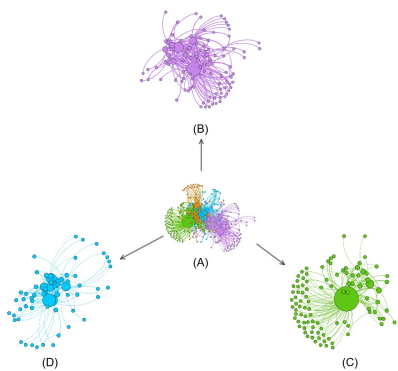

Figure 5. The Military TV channel's co-commenter network (A) and the three largest commenter mobs (B, C, and D).

To recap, we observed similarities among the 20 YouTube channels with false views about the U.S. Military. These similarities were captured by analyzing the network structural features of co-commenter networks on these channels and applying k-means and hierarchical clustering methods. PCA was used to reduce the dimensionality. The similarities in commenters' behaviors suggested a high level of collusion (or coordination) among the channels. Three most suspicious channels were identified where commenter mobs were detected. Furthermore, cross-channel commenter mobs were identified on two of the three most suspicious channels, meaning the commenter mob moved from one channel to the other.

VI. CONCLUSION AND FUTURE WORK

In this study, we present a methodology to identify suspicious commenter behaviors by analyzing co-commenter networks, in particular on YouTube. We developed a network analysis based approach that leverages 20 network structural features to assess suspicious behaviors in commenter networks, which enables channel-level characterization. Furthermore, the proposed methodology helps identifying behavioral similarities among the channels based on commenter network structures, which allows assessment of varying degrees of collusion (or coordination). We evaluated the proposed methodology on a set of 20 YouTube channels that post videos containing untrue and unflattering views of the U.S. Military. Our methodology helped identify the top 3 most suspicious channels. A deeper analysis of these three channels revealed false narratives being pushed in their videos. The methodology also revealed clusters of channels with similar commenter behavioral profiles that range from highly suspicious to semi-suspicious.

To advance our understanding of suspicious commenter behavior on online platforms like YouTube, future research should analyze the motivations, interests, and power of the users willing to participate in suspicious activities. Further-more, there is a requirement for an extended research study to explore behavior patterns associated with suspicious activities, such as topic modeling, sentiment scoring, and toxicity analysis. Another future study will investigate the potential impact of suspicious commenter behavior on the spread of misinformation and the integrity of online discussions. Utilizing the focal structure analysis models [22] could help to observe mob activities on suspicious YouTube channels, with the importance of identifying any commenters' collusion or coordination.

## Acknowledgment

This research is funded in part by the U.S. National Science Foundation (OIA-1946391, OIA-1920920, IIS-1636933, ACI-1429160, and IIS-1110868), U.S. Office of the Under Secretary of Defense for Research and Engineering (FA9550-22-1-0332), U.S. Office of Naval Research (N00014-10-1-0091, N00014-14-1-0489, N00014-15-P-1187, N00014-16-1-2016, N00014-16-1-2412, N00014-17-1-2675, N00014-17-1-2605, N68335-19-C-0359, N00014-19-1-2336, N68335-20-C-0540, N00014-21-1-2121, N00014-21-1-2765, N00014-22-1-2318), U.S. Air Force Research Laboratory, U.S. Army Research Office (W911NF-20-1-0262, W911NF-16-1-0189, W911NF-23-1-0011), U.S. Defense Advanced Research Projects Agency (W31P4Q-17-C-0059), Arkansas Research Alliance, the Jerry L. Maulden/Entergy Endowment at the University of Arkansas at Little Rock, and the Australian Department of Defense Strategic Policy Grants Program (SPGP) (award number: 2020-106-094). Any opinions, findings, and conclusions or recommendations expressed in this material are those of the authors and do not necessarily reflect the views of the funding organizations. The researchers gratefully acknowledge the support.